\documentclass[conference]{IEEEtran}
\IEEEoverridecommandlockouts
\usepackage{cite}
\usepackage{amsmath,amssymb,amsfonts}
\usepackage{algorithmic}
\usepackage{graphicx}
\usepackage{textcomp}
\usepackage{xcolor}

\usepackage{afterpage}
\usepackage{multirow}
\usepackage{wrapfig}
\usepackage{graphicx}
\usepackage{hyperref}

\usepackage{booktabs} 

\def\BibTeX{{\rm B\kern-.05em{\sc i\kern-.025em b}\kern-.08em
    T\kern-.1667em\lower.7ex\hbox{E}\kern-.125emX}}
\begin{document}

\title{Latent CLAP Loss for Better Foley Sound Synthesis
}

\makeatletter
\newcommand{\linebreakand}{%
  \end{@IEEEauthorhalign}
  \hfill\mbox{}\par
  \mbox{}\hfill\begin{@IEEEauthorhalign}
}

\makeatother

\author{\IEEEauthorblockN{Tornike Karchkhadze$^{\dag}$}
\IEEEauthorblockA{\textit{University of California San Diego} \\
San Diego, USA \\
tkarchkhadze@ucsd.edu}
\and
\IEEEauthorblockN{Hassan Salami Kavaki$^{\dag}$}
\IEEEauthorblockA{\textit{The Graduate Center, CUNY} \\
New York, USA \\
hsalami@gradcenter.cuny.edu}
\and
\IEEEauthorblockN{Mohammad Rasool Izadi}
\IEEEauthorblockA{\textit{Bose Corp.} \\
Framingham, USA \\
russell\_izadi@bose.com
}
\and
\IEEEauthorblockN{Bryce Irvin}
\IEEEauthorblockA{\textit{Bose Corp.} \\
Framingham, USA \\
bryce\_irvin@bose.com}
\linebreakand
\IEEEauthorblockN{Mikolaj Kegler}
\IEEEauthorblockA{\textit{Bose Corp.} \\
Framingham, USA \\
mikolaj\_kegler@bose.com}
\and
\IEEEauthorblockN{Ari Hertz}
\IEEEauthorblockA{\textit{Bose Corp.} \\
Framingham, USA \\
ari\_hertz@bose.com}
\and
\IEEEauthorblockN{Shuo Zhang}
\IEEEauthorblockA{\textit{Bose Corp.} \\
Framingham, USA \\
shuo\_zhang@bose.com}
\and
\IEEEauthorblockN{Marko Stamenovic}
\IEEEauthorblockA{\textit{Bose Corp.} \\
Framingham, USA \\
marko\_stamenovic@bose.com}
}


\maketitle

\def\thefootnote{\dag}
\footnotetext{Authors with equal contribution. This work was performed during an internship at Bose Corporation.}



\begin{abstract}
Foley sound generation, the art of creating audio for multimedia, has recently seen notable advancements through text-conditioned latent diffusion models. These systems use multimodal text-audio representation models, such as Contrastive Language-Audio Pretraining (CLAP), whose objective is to map corresponding audio and text prompts into a joint embedding space. AudioLDM, a text-to-audio model, was the winner of the DCASE2023 task 7 Foley sound synthesis challenge. The winning system fine-tuned the model for specific audio classes and applied a post-filtering method using CLAP similarity scores between output audio and input text at inference time, requiring the generation of extra samples, thus reducing data generation efficiency. We introduce a new loss term to enhance Foley sound generation in AudioLDM without post-filtering. This loss term uses a new module based on the CLAP model—Latent CLAP encoder—to align the latent diffusion output with real audio in a shared CLAP embedding space. Our experiments demonstrate that our method effectively reduces the Fréchet Audio Distance (FAD) score of the generated audio and eliminates the need for post-filtering, thus enhancing generation efficiency.

\end{abstract}


\section{Introduction}

In recent years, deep learning models have made tremendous advances in the domain of sound generation~\cite{pmlr-v202-liu23f, liu2023audioldm, Yang2022, kreuk2023audiogen, huang2023makeanaudio}. In light of these advancements, user-controlled neural audio synthesis has the potential to revolutionize numerous domains, including Foley sound synthesis~\cite{Chen2017, Iashin2021}, the art of creating or reproducing everyday sound effects that are added to film, video, and other media in post-production. This innovation could impact a variety of fields that utilize Foley sound by automating the labor-intensive manual sound design.

Diffusion models~\cite{ho2020denoising} have gained significant attention for their ability to learn complex distributions, which makes them well-suited for data types such as audio. By integrating the latent diffusion model (LDM)~\cite{rombach2022high} for generation and the contrastive language-audio pretraining (CLAP)~\cite{wu2023largescale} model as a text encoder, AudioLDM~\cite{pmlr-v202-liu23f} stands as one of the current state-of-the-art text-to-audio generation systems, demonstrating strong performance in Foley sound generation tasks. Models built upon AudioLDM~\cite{yuan2023textdriven,yuan2023latent} emerged as the winner of the DCASE2023 Task 7 Foley Sound Synthesis Challenge~\cite{Choi_arXiv2023_01}. The winning model used AudioLDM, fine-tuned on a set of sound classes, alongside a post-filtering technique that generated superfluous samples and sub-selected the best among them using a heuristic. Although post-filtering improved the output quality, it significantly compromised the efficiency of generation.

We propose augmenting the AudioLDM framework by introducing a Latent CLAP loss, which is incorporated through a Latent CLAP encoder module. This module is integrated into the fine-tuning process with an objective to increase the similarity in the CLAP embedding space between the generated and real audio at training time, thus enhancing the quality of the generated samples.

Our experiments show that the proposed method not only improves Fréchet Audio Distance (FAD)~\cite{kilgour2019frechet}, an objective generative audio metric,  but also alleviates the need for post-filtering, thereby enhancing the quality of generated sounds while greatly improving the efficiency of the audio generation process. Additionally, a perceptual study involving subjective evaluations from a cohort of human listeners confirm our objective metrics, indicating strong preferences towards the data generated using the proposed model, as compared to AudioLDM fine-tuned on the DCASE2023 Task 7 data.

\section{Background} \label{section_background}

\subsection{AudioLDM System Overview}
\label{sec:audioldm-background}

AudioLDM comprises a text-audio encoder, a generator, an autoencoder, and a vocoder. CLAP serves as the text-audio encoder, mapping audio and text into a shared embedding space. This is succeeded by an LDM that serves as the main generator of the system. To train on limited computational resources while retaining generation quality, the LDM operates within the latent space of a Variational Autoencoder (VAE)~\cite{kingma2022autoencoding}. This VAE is pretrained to compress and reconstruct Mel-spectrograms. Finally, a HiFi-GAN~\cite{kong2020hifi} vocoder is used to synthesize the audio output from the generated Mel-spectrograms.

Each component—the CLAP encoder, VAE, and HiFi-GAN vocoder—is pretrained independently and frozen within the overall system, for the subsequent training of the LDM. AudioLDM overcomes a primary challenge in the field of audio machine learning, the need for vast, high-quality audio-to-text datasets, by directly extracting embeddings from audio using the CLAP model. Additionally, AudioLDM utilizes classifier-free guidance (CFG)~\cite{ho2022classifierfree}, which allows for the adjustment of adherence to conditioning during inference through CFG weighting. 

\subsection{Foley Sound Synthesis Task and Baseline Models}
\label{sec:baselines}

This study follows the challenge set out by DCASE2023, specifically Task 7 on Foley Sound Synthesis~\cite{Choi_arXiv2023_01}. The challenge was set up as a category-to-sound generation with seven distinct categories: dog bark, footstep, gunshot, keyboard, moving motor vehicle, rain, and sneeze/cough. The challenge organizers supplied a hand-curated development dataset containing the above mentioned 7 categories. 

For our baseline comparisons, we select the winning model from the DCASE2023 Challenge Task 7 on Foley Sound Synthesis~\cite{yuan2023textdriven}, and its subsequent modification~\cite{yuan2023latent}. Both baselines employ a pretrained AudioLDM, fine-tuned on the text-to-audio pairs from the challenge dataset. In this process, class labels are transformed into descriptive sentences, which are then fed into the text branch of the CLAP model. To extract better semantic embeddings for each class, ~\cite{yuan2023latent} used an additional tuning layer between the CLAP and LDM modules. Both models applied a post-filtering technique to the generated audio samples. The post-filtering process involved calculating the cosine similarity between CLAP embeddings of the generated audio and the target text. Audio samples achieving similarity scores beyond a predefined threshold were then selected for the final output pool.
In both studies, the post-filtering thresholds, as well as the CFG weights~(see~\ref{sec:audioldm-background}) were determined empirically for each sound class.

\section{Method} \label{section_method}
\label{sec:method}

Our system, depicted in Fig.~\ref{fig1}, is a modification of AudioLDM~\cite{liu2023audioldm} with the addition of a novel Latent CLAP encoder for enhanced loss computation.

\begin{figure}[t]
  \centering
  \includegraphics[width=.95\linewidth]{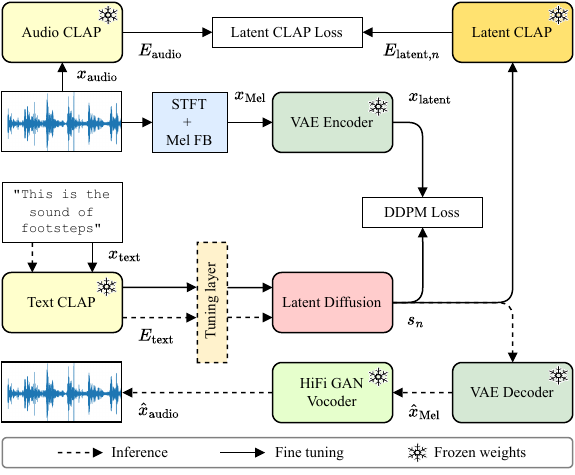}
  \setlength{\abovecaptionskip}{-1pt}
  \caption{Overview of the overall system for Foley audio generation.}
  \vspace{-9pt}
  \label{fig1}
\end{figure}

\subsection{CLAP Encoder}

The CLAP model comprises two parallel, jointly-trained transformer-based encoders: one for text and one for audio. The text encoder in the CLAP model converts the text input \( x_{\text{text}} \) into an embedding vector denoted by \( E_{\text{text}} \). Similarly, the audio encoder converts the audio input \( x_{\text{audio}} \) into \( E_{\text{audio}} \) that matches the dimensions of \( E_{\text{text}} \). These two encoders are trained with a contrastive loss, resulting in an aligned latent space of the same dimensionality. A comparative analysis ~\cite{wu2023largescale} shows that the best architecture for the CLAP audio encoder is HTSAT~\cite {9746312}, and the best architecture for the text encoder is the RoBERTa~\cite{DBLP:journals/corr/abs-1907-11692} language model. In our work, we utilize the publicly released model \footnote{https://github.com/LAION-AI/CLAP} from~\cite{wu2023largescale} pretrained on AudioCaps~\cite{kim-etal-2019-audiocaps}, Clotho~\cite{drossos2019clotho}, LAION-Audio-630K~\cite{wu2023largescale}, and Audioset~\cite{45857} datasets. 

Similarly to the baseline model~\cite{yuan2023latent}, we add a fully connected tuning layer after the text CLAP encoder (see Fig.~\ref{fig1}). We evaluate the system with and without this layer.

\subsection{LDM Generator}

The LDM generator, operating in latent space, when conditioned on the text embedding \( E_{\text{text}} \), produces a compressed representation of Mel-spectrogram \( s_n \in \mathbb{R}^{C \times \frac{T}{r} \times \frac{F}{r}} \), where \( r \) shows the compression ratio, \( C \) denotes the number of channels,  \( T \) and \( F \) represent the time and frequency dimensions of Mel-spectrogram respectively. The variable \( n \in [1, \ldots, N] \) in \( s_n \) represents the step number in the diffusion model's forward or reverse steps. During the forward phase of the diffusion process, the original clean compressed representation of the Mel-spectrogram \( s_0 \) is gradually turned into an isotropic Gaussian noise \( s_N \) with distribution \( \mathcal{N}(0, I) \) over \( N \) steps, by incrementally adding Gaussian noise \( \epsilon \) at each step. Conversely, in the reverse phase of the diffusion process, the clean representation \( s_0 \) is incrementally reconstructed through a denoising process, where at each time step \( n \), the model predicts the injected noise \( \epsilon \) to reconstruct \( s_{n-1} \) from \( s_n \).

The training objective of the model is to minimize the mean square error (MSE) between the predicted noise \( \epsilon_{\theta}\) and the actual Gaussian noise \( \epsilon \), following the classic Denoising Diffusion Probabilistic Model's (DDPM)~\cite{ho2020denoising} loss approach:
\begin{align}
{L_{DDPM}}({\theta})= \mathbb{E}_{s_0, \epsilon, n} \|\epsilon - \epsilon_{\theta}(s_n,n, E_{\text{text}}) \|^2,
\label{eq1}
\end{align}
where $\theta$ corresponds to the parameters of the LDM model.

\subsection{Variational Autoencoder and HiFi-GAN Vocoder}

We employ a VAE model~\cite{Iashin2021} to bridge the Mel-spectrogram domain with the latent space used by the LDM. 
The VAE encoder projects Mel-spectrograms into the latent space and the VAE decoder reconstructs the Mel-spectrograms from compressed representations. During training, the input audio \( x_{\text{audio}} \) is first converted into a Mel-spectrogram \( x_{\text{Mel}} \) as shown in Fig. \ref{fig1}. Then, the VAE encoder maps \( x_{\text{Mel}} \) to a latent compressed representation \( x_{\text{latent}} \) \( \in \mathbb{R}^{C \times \frac{T}{r} \times \frac{F}{r}} \), which is then compared with the LDM's output \( s_n \). During inference, the VAE decoder transforms the denoised compressed representation output of LDM, \( s_0 \), into the Mel-spectrogram \( \hat{x}_{\text{Mel}} \). The HiFi-GAN vocoder subsequently maps \( \hat{x}_{\text{Mel}} \) into an audio waveform \( \hat{x}_{\text{audio}} \).

\subsection{Latent CLAP Encoder}


To improve the model performance without the need for post-filtering, we attempted to include CLAP similarity score into the loss function, aiming to maximize similarity during training. Unfortunately, this end-to-end training setup required extensive GPU resources and proved to be ineffective. To overcome this limitation, we developed a new module that maps the VAE latent space, in which LDM operates, to the CLAP embedding space. This module, which we refer to as the Latent CLAP encoder, utilizes the Pretrained Audio Neural Network (PANN) architecture~\cite{kong2020panns}, a popular approach for tasks such as audio classification. Concretely, we use the `PANN-10' design, with modified input and output layers which process compressed audio representations -- \( x_{\text{latent}} \) during its own training and \( s_n \) during fine-tuning of the AudioLDM system -- and predicts the corresponding CLAP embedding \(E_{\text{latent}, n} \).

As illustrated in Fig. \ref{fig2}, we trained the Latent CLAP encoder $\phi$ in conjunction with the frozen audio branch of the pretrained CLAP model using MSE objective to match the embedding spaces of the two encoders:
\begin{equation}
    L(\phi) = \mathbb{E}_{x_{\text{latent}}} \|E_{\text{audio}} - E_{\text{latent}}\|^2.
    \vspace{-1pt}
\end{equation}
The audio inputs \( x_{\text{audio}} \) are first transformed into Mel-spectrograms \( x_{\text{Mel}} \) and then compressed into a latent representation \( x_{\text{latent}} \) by a pretrained, frozen VAE encoder before being fed into Latent CLAP. The dataset employed for training was WaveCaps~\cite{mei2023wavcaps}, a collection of around 400K Chat-GPT labeled audio clips. Apart from the labels that were not used in our training process, the WaveCaps compilation includes popular audio datasets such as Audioset, BBC, Freesound, and Soundbible, sharing substantial commonalities with the data that was used to train the original CLAP model. Our rationale for using a similar dataset was to align the performance of the new Latent CLAP encoder as closely as possible with the existing branches of CLAP.

\begin{table*}[!t]
    \caption{Evaluation: FAD Scores (lower is better, $\downarrow$).} 
    \centering
    \resizebox{\textwidth}{!}{
        \begin{tabular}{c|ccccccc|cc}
            \toprule
            System & Dog Bark & Footstep & Gun Shot & Keyboard & Motor Vehicle & Rain & Sneeze/Cough & Average & Std. \\
            \midrule
            Real Audio & 3.19 & 5.51 & 5.02 & 3.42 & 6.24 & 4.46 & 4.43 & 4.61 & 1.09 \\
            \midrule
            LDM~\cite{yuan2023textdriven} & 5.69 & 10.24 & 7.25 & 5.11 & 19.77 & 9.75 & 3.38 & 8.74 & 5.45 \\
            LDM+Tuning~\cite{yuan2023latent} & 6.58 & 10.31 & 5.07 & 4.7 & 11.57 & 9.59 & 4.5 & 7.47 & 2.95 \\
            LDM+Latent & \textbf{4.54} & \textbf{7.08} & 5.6 & 5.31 & 16.14 & \textbf{7.84} & \textbf{3.16} & 7.1 & 4.28 \\
            LDM+Tuning+Latent & 5.91 & 9.9 & \textbf{4.51} & \textbf{4.46} & \textbf{7.3} & 8.99 & 4.55 & \textbf{6.52} & \textbf{2.26} \\
            \midrule
            LDM+filter~\cite{yuan2023textdriven} & \textbf{4.11} & 9.47 & 5.2 & \textbf{4.48} & 23.45 & 9.56 & 3.4 & 8.52 & 7.05 \\
            LDM+Tuning+filter~\cite{yuan2023latent} & 5.64 & 9.91 & 4.56 & 4.93 & 12.54 & 8.86 & 4.33 & 7.25 & 3.2 \\        
            LDM+Latent+filter & 4.46 & \textbf{8.24} & 5.27 & 4.71 & 10.91 & \textbf{8.83} & \textbf{2.87} & \textbf{6.47} & 2.88 \\
            LDM+Tuning+Latent+filter & 6.24 & 9.01 & \textbf{4.4} & 4.86 & \textbf{10.7} & 9.94 & 4.24 & 7.05 & \textbf{2.76} \\
            \bottomrule
        \end{tabular}
    }
    \label{tab: FAD and Similarity Scores}
    \vspace{-5pt}
\end{table*}

As shown in Fig. \ref{fig1}, we introduce the Latent CLAP loss to our Foley generation system at the fine-tuning stage, aiming to enhance the similarity between LDM's latent audio output and the actual ground truth audio, facilitating improvements in sample quality. The objective of the loss function in the training process is minimizing the MSE between the CLAP embedding vector outputs of real audio and generated latent representations, as follows:
\begin{align}
{L_{LCLAP}}({\theta}) &= \mathbb{E}_n \Bigl [ w(n) \cdot \| E_{\text{audio}} - E_{\text{latent}, n} \|^2 \Bigl ],
\label{eq3}
\end{align}
where the embedding vectors \( E_{\text{audio}} \) and \( E_{\text{latent}, n} \) represent the outputs of the CLAP audio branch and latent CLAP encoder, respectively. The audio branch processes the incoming audio \(x_{\text{audio}}\), while the Latent CLAP encoder operates on the noisy, compressed representation \(s_n\), which is the output of the LDM model at step \(n\). The \( w(n) \) in the equation is a \( n \)-dependent weighting function that defines the contribution of higher \( n \) (noisier) and lower \( n \) (cleaner) samples in the loss calculation, addressing the observed ineffectiveness of the method when treating noisy and less noisy samples uniformly.

The Foley synthesis model's total loss combines the conventional DDPM loss with the newly introduced Latent CLAP loss with weight \( \lambda \) as follows:
\begin{align}
L({\theta}) = {L_{DDPM}}({\theta}) + \lambda {L_{LCLAP}}({\theta})
\label{eq4}
\end{align}

\begin{figure}[t]
  \centering
    \includegraphics[width=0.85\linewidth]{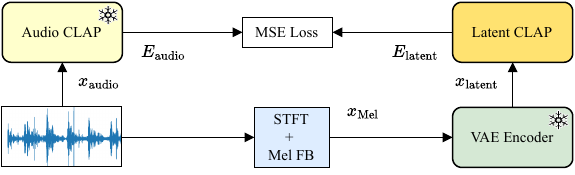}
    \setlength{\abovecaptionskip}{-1pt}
    \caption{ Latent CLAP encoder pretraining.}
  \vspace{-9pt}
  \label{fig2}
\end{figure}

\section{Experimental Setup} \label{section_experimental_setup}

\subsection{Data and Metrics}

Our experiments utilized the DCASE2023 Task 7 challenge dataset, comprising 6.1 hours of audio, distributed across seven given categories, with each containing 600 to 800 clips of 4 seconds long. We pre-processed the data by resampling it from 22 kHz to 16 kHz and employing a repeating method to extend the clips from 4 to 10.24 seconds in length. Using a window length of 1024 and a hop size of 160 samples, the audio clips were transformed into Mel-spectrograms with dimensions \( F \times T = 64 \times 1024 \), representing Mel-frequency bins and time frames respectively. Text prompts were transformed from class labels to sentences using the phrase “\texttt{This is a sound of <label>}.” We adopted the FAD metric for evaluation, utilizing the means and standard deviations of VGGish embeddings provided by the challenge organizers as the reference distribution for the evaluation set.

\subsection{Parameter Settings}

As a starting point for our fine-tuning process, we took a pretrained \textit{audioldm-m-full} variant of the AudioLDM model from publicly released checkpoints\footnote{\href{https://github.com/haoheliu/AudioLDM}{github.com/haoheliu/AudioLDM}}. When fine-tuning with the DCASE2023 development dataset, all parts of the model were frozen except for the parameters of the LDM and the tuning layer as demonstrated in Fig. \ref{fig1}. Our setup uses a VAE with a compression ratio \( r = 4\), encoding Mel-spectrograms of size \( F\times T = 64 \times 1024 \) into latent vectors of size \( C \times \frac{F}{r} \times \frac{T}{r} = 8 \times 16 \times 256 \) in channel, frequency and time dimensions, respectively. The embedding sizes for audio, text, and latent representations from the CLAP encoder are 512. During fine-tuning, we use the Adam optimizer with a learning rate of \( 3 \times 10^{-6} \) for up to 500 epochs. The number of LDM denoising steps is set to  \( N = 1000 \) during training and \( N = 200 \) during inference. To evaluate model performance, we generate 100 clips per class and compute their FAD with respect to the evaluation set.

In our experiments, we standardized the CFG weight~(see~\ref{sec:audioldm-background}) to 2.0 for all categories, unlike the baselines~\cite{yuan2023textdriven,yuan2023latent} in which it varied between 1.5 and 2.5. Also in contrast to the baselines, which used varying AudioLDM models for different classes, we only used \textit{audioldm-m-full}, with a goal of analyzing Latent CLAP's impact rather than minimizing FAD scores. For post-filtering, we used the same CLAP similarity threshold as in the baselines~(see~\ref{sec:baselines}). Namely 0.2 for all classes, except for keyboard (0.15) and motor vehicle (0.75).

For Latent CLAP loss in \eqref{eq3}, we set the denoising-step-dependent weighting function as $w(n) = 10^{-\frac{n}{200}}$, where $n$ is the denoising step in the latent diffusion process. The function is an exponent decay attenuating contributions of overly noisy samples in the training above $N= 200$ step, which is also the number of steps used at the inference stage. For the weighting coefficient $\lambda$ balancing the DDPM and Latent CLAP loss contributions in \eqref{eq4}, we found that a setting of \( \lambda = 2000 \) provided the best results for the model using latent CLAP loss, whereas \( \lambda = 1000 \) yielded the best results for the model utilizing both latent CLAP loss and the tuning layer.

\section{Results} \label{section_results}

\subsection{Objective Metrics}

We evaluated our models, which incorporate Latent CLAP loss, both with and without the inclusion of the tuning layer, against baselines~\cite{yuan2023latent,yuan2023textdriven}. The performance of our proposed method on the DCASE2023 Task 7 validation set is reported in Table \ref{tab: FAD and Similarity Scores}. As shown in rows 2-5, models using Latent CLAP loss in both settings outperform the baselines in terms of FAD score (1.64 and 0.95 avg. FAD improvement, respectively), indicating substantially better correspondence of the generated set with the evaluation set in the matter of objective metrics.

We observed that integration of Latent CLAP loss showed minimal effect on the similarity scores between generated audio and text embeddings ($.65 \rightarrow .66$ with the tuning layer and $.30 \rightarrow .31$ without the tuning layer). Our results demonstrate that the addition of latent CLAP enhances the output audio quality without altering the alignment between the text prompts and generated audio. This outcome is consistent with our emphasis on improving audio fidelity, as opposed to increasing the similarity between text and audio, which is the underlying assumption of the baseline's use of post-filtering.

Furthermore, we observed that audio and text embedding similarity score-dependent post-filtering is ineffective for our model (Table \ref{tab: FAD and Similarity Scores}, lower section). In contrast to the baseline models, where post-filtering has a noticeable effect, our method, particularly combining the tuning layer and Latent CLAP loss, shows that post-filtering not only becomes superfluous and can also slightly increase the FAD score.

The elimination of the post-filtering stage provided significant data generation efficiency improvements. Here, using NVIDIA GeForce RTX 3090 generating 100 10-seconds-long clips, the elimination of post-filtering increases the generation speed by an order of magnitude. 

\begin{figure}[tb]
  \centering
  \includegraphics[width=\linewidth]{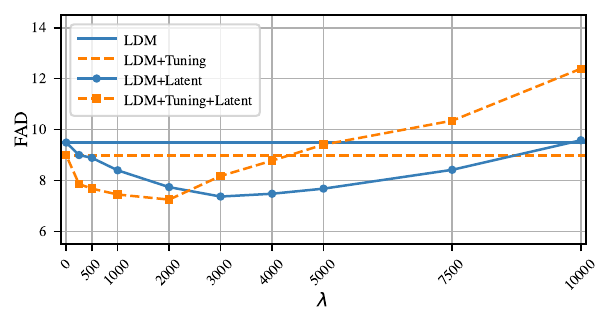}
  \setlength{\abovecaptionskip}{-20pt}
  \caption{Impact of Latent CLAP Loss Weight on FAD Score. Note: these FAD scores were obtained for the models trained for only 100 epochs, while Table~\ref{tab: FAD and Similarity Scores} presents the evaluation of the models trained for 500 epochs. This study aims to explore the relative performance changes as a functions of $\lambda$.}
  \vspace{-15pt}
  \label{FAD_VS_Lambda}
\end{figure}

\subsection{Latent CLAP Loss Weight Analysis}  

To test the robustness of Latent CLAP loss to hyperparameter selections, we varied the weighting parameter \( \lambda \) from \eqref{eq4} and calculated the average FAD scores across all classes, comparing models with and without the tuning layer trained over 100 epochs (note that the models in Table~\ref{tab: FAD and Similarity Scores} were trained for 500 epochs). 
Results presented in Fig.~\ref{FAD_VS_Lambda} illustrate that the increase in the $\lambda$ gradually improves the avg. FAD score until the optimal point is reached, beyond which the performance starts to decrease. This indicates that the latent CLAP loss is consistently contributing to the overall performance without unexpected $\lambda$-dependent effects.

\subsection{Subjective Evaluation}

Human perception of audio quality and class relevance can be highly subjective and nonlinear, often correlating poorly with objective metrics~\cite{vinay2022evaluating}. Human ratings are still considered the gold standard in audio quality evaluation. Thus to better ground our work, the audio generated by the proposed method was compared to those from the baseline model involving the tuning layer (LDM + Tuning)~\cite{yuan2023latent} and ground truth audio samples in a series of 7 online surveys. Each survey focused on a specific sound class, included 15 audio samples from each of the three models and was completed by 40 listeners. Respondents rated audio clip for their quality and class relevance on a scale from 0 to 10. Our survey's audio samples and analyses are publicly available\footnote{ \url{https://github.com/karchkha/Latent-CLAP-subjective-evaluation}}.

As reported in Table~\ref{tab_subjective_evaluation}, the average subjective ratings reveal that the Latent CLAP loss model outscored the baseline model and ground truth samples in both audio quality and category fitness. A mixed-design ANOVA showed that the main effect of model was significant for both audio quality ($F$(1.9,509.6) = 115.3, $p$ $<$ .001, $\eta_{p}^{2}$ = .3) and category fit ($F$(1.9,522) = 156.6, $p$ $<$ .001, $\eta_{p}^{2}$ = .36), while post-hoc paired t-tests with Bonferroni corrections further support the finding that the proposed model was rated significantly higher than both the baseline model and ground truth samples across both items ($p$ $<$ .001 for all comparisons). 

The lower ratings for the ground truth samples is unexpected and could be attributed to the fact that real-world recordings often contain extraneous noises or recording artifacts. Our generated audio presents cleaner sounds with features that are more distinctly aligned with the target class, offering a potentially clearer representation of the intended sound event, which may contribute to higher scores in human evaluations.

\begin{table}[t]
\caption{Subjective evaluation of Audio Quality and class relevance}
    \centering
    \begin{tabular}{lcc}
    \toprule
    Model & Audio quality & Class relevance \\
    \midrule
    Ground Truth & 5.55 & 6.03 \\
    LDM + Tuning~\cite{yuan2023latent} & 6.00 & 6.55 \\
    LDM + Tuning + Latent & \textbf{6.40} & \textbf{7.03} \\
    \bottomrule
    \end{tabular}
    \label{tab_subjective_evaluation}
    \vspace{-10pt}
\end{table}

\section{Conclusion}

This work introduces an improved AudioLDM model for Foley sound generation with the addition of the Latent CLAP loss. 
The concept of the proposed method is to integrate the post-filtering process within the model to enhance audio quality while maintaining inference efficiency. As indicated by both objective (FAD score) and subjective evaluations (listening study), the proposed model generates higher quality audio in comparison to baselines. This approach is not limited to the closed-set Foley generation problem and can be applied to general-purpose text-to-audio models without any further modifications. Future research will explore these applications further, potentially enhancing the fidelity of generative audio models.

\bibliographystyle{IEEEtran}
\bibliography{IEEEabrv,refs}


\end{document}